\documentclass{dcdsi} % Use the dcdsc.cls file to compile your paper

\def\ArticleType{Article}  % Default option

\usepackage{bbding}

\usepackage{bm}	
\usepackage{hyperref}
\usepackage{cleveref}
\crefname{section}{Section}{Sections}
\crefname{subsection}{Section}{Sections}

\usepackage{booktabs}

\allowdisplaybreaks

\def\currentvolume{xx}
\def\currentnumber{xx}
\def\currentyear{20xx}
\def\articlenumber{Axxxxxx} % Default option
\def\DOI{10.3934/dcdsi.\articlenumber}

\theoremstyle{plain}

\theoremstyle{definition}

\title{Fold of a bifurcation solution from the figure-eight choreography in the three-body problem}

\author[a]{\scshape Hiroshi Fukuda\textsuperscript{\href{mailto:fukuda@kitasato-u.ac.jp}{\Envelope}}}
\author[b]{\scshape Hiroshi Ozaki\textsuperscript{\href{mailto:ozaki@tokai.ac.jp}{\Envelope}}}
\begin{document}

\thispagestyle{cover}
\FirstpageLayout

\maketitle

% Enter authors' affiliations
%\FirstpageAffiliations{%
%	\textsuperscript{a}College of liberal arts and sciences, Kitasato University, Kitasato 1-15-1, 
%	Minami-ku, Sagamihara, Kanagawa 252-0373, Japan.
%	\textsuperscript{b}STEM Education Center, Tokai University, 4-1-1, Kitakaname, Hiratsuka, Kanagawa 259-1292, Japan.
%%	\textsuperscript{c}Affiliation Three.
%}
\FirstpageAffiliations{%
	\textsuperscript{a}College of Liberal Arts and Sciences, Kitasato University, Kitasato 1-15-1, 
	Minami-ku, Sagamihara, Kanagawa 252-0373, Japan.
	\textsuperscript{b}Independent Researcher, Yokohama, Kanagawa 234-0054, Japan.
}

%	Your abstract here (maximum 200 words).
\begin{abstract}
	In the figure-eight choreography of the classical three-body problem, bifurcated solution branches may undergo a fold transition immediately after their emergence, which appears as a cusp in the action. We show that some of this phenomenon is governed by a universal mechanism in the fourth-order Lyapunov--Schmidt reduced action for a two-dimensional representation with three-fold symmetry. We derive an explicit critical condition for the occurrence of a fold in terms of the ratio of the third- and fourth-order coefficients. We illustrate this mechanism with four numerical examples: three for a Lennard-Jones-type potential and one for a homogeneous potential.
\end{abstract}
% Please provide a minimum of 5 keywords or phrases.
%\keywords{bifurcation, three-body, cusp, figure-eight, choreography, Lyapunov-Schmidt}
%\keywords{Three-body problem, Figure-eight choreography, Bifurcation analysis, Inhomogeneous potentials, Lyapunov-Schmidt reduction}
%%%%%%%%%%%%%%%%%%%%%%%%%%%%%%%%%%%%%%%%%%%%%%%%%%%%%%%%%%%%%%%%%%
%      5. Significance/Author contributions/Competing interests
%         corresponding author/Mathematics Subject Classification
%%%%%%%%%%%%%%%%%%%%%%%%%%%%%%%%%%%%%%%%%%%%%%%%%%%%%%%%%%%%%%%%%%

\ReceivedDate{xxxx xx, 20xx} %{Month date, year}
\AcceptedDate{xxxx xx, 20xx} %{Month date, year}
\PublishedDate{xxxx xx, 20xx} % The information below will be filled in by AIMS production staff.

%\Significance{
%	Authors are required to submit a significance statement of 50–120 words that describes the importance of their research. The statement should be written at an understandable level for an undergraduate-educated scientist outside the authors' field of specialization and should clearly explain the broader relevance and impact of the work to a general scientific audience.
%}
\Significance{Periodic motions in few-body systems can change their structure when a system parameter is varied, but the geometric mechanisms underlying such changes are often difficult to identify in the full equations of motion. We show that a broad class of symmetry-breaking bifurcations in the figure-eight three-body choreography possesses a universal local structure: bifurcated solution branches can collide in a fold, producing a cusp in the action. This structure is derived analytically from a two-dimensional reduced action with three-fold symmetry and is confirmed numerically for four distinct bifurcations under both inhomogeneous and homogeneous potentials. Our results provide a general framework for recognizing fold phenomena in symmetric few-body dynamics.}

\FirstPageRightBlockBottom{%
	\textbf{Author contributions}:
	Hiroshi Fukuda designed and performed research;
	Hiroshi Ozaki scientific consultation;\\
	\textbf{Competing interests}: The authors declare no competing interests.\\
	%
	% \textsuperscript{1}\textbf{Co-first authors}\\
	%
	\CorrespondingAuthor{fukuda@kitasato-u.ac.jp}\\
	\textbf{Handling Editor}: Associated Editor\\
	%
	% 2020 Mathematics Subject Classification (MSC)
	\Subjclass{Primary: 70F07, 37G15; Secondary: 70F15.}
}
\FirstPageRightBlock

%%%%%%%%%%%%%%%%%%%%%%%%%%%%%%%%%%%%%%%%%%%%%%%%%%%%%%
%                   6. BODY
%%%%%%%%%%%%%%%%%%%%%%%%%%%%%%%%%%%%%%%%%%%%%%%%%%%%%%

% Only the first word and proper nouns of section titles should be capitalized.

\section{Introduction}
The figure-eight choreography in the classical three-body problem is a remarkable periodic orbit in which three equal masses chase one another along a fixed figure-eight-shaped curve in the plane \cite{moore,Chenciner}. As system parameters are varied, this choreography can undergo bifurcations that alter the structure of solutions. 
Many such bifurcations are equivariant ones \cite{CL}, in which a branch of solutions may spontaneously break part of the full spatiotemporal symmetry described by a crystallographic point group \cite{Chenciner2, fukuda2025}.

For inhomogeneous potentials such as the Lennard--Jones-type (LJ) potential \cite{sbano}, bifurcations can be induced naturally by varying the period $T$ of the motion without changing the Lagrangian \cite{fukuda2019}. On the other hand, for homogeneous potentials, bifurcations are typically induced by varying either the power of the potential \cite{fukuda2019} or one of the masses of the bodies \cite{galan2002, doedel2003}.

We note that some bifurcated branches undergo a fold immediately after their emergence \cite{fukuda2019, fukuda2025}. In this paper, this fold structure is analyzed by applying the Lyapunov--Schmidt (LS) reduction to the action functional, yielding a two-dimensional reduced action up to the fourth order.

Throughout this study, we assume that the symmetry of the Lagrangian remains unchanged as the bifurcation parameter is varied. Varying the period of the motion or the power of a homogeneous potential preserves the symmetry of the Lagrangian. In contrast, varying one of the masses explicitly breaks the symmetry associated with the equal-mass configuration and is therefore outside the scope of the present analysis.

In \cref{sec:def}, the bifurcation and the LS reduction of the action functional to a two-dimensional representation are reviewed according to our previous work \cite{fukuda2025} to provide the setting for the new analysis developed in the following subsections. In \cref{sec:sol}, the critical-point equations for the LS reduced action with three-fold symmetry are solved up to the fourth order.
We demonstrate that the bifurcated solutions undergo a fold under a specific condition determined by the third- and fourth-order expansion coefficients. In \cref{relative_action}, the cusp structure in the relative action of a bifurcated solution relative to the original solution is derived. In \cref{sec:top}, the two-dimensional LS reduced action is visualized using three-dimensional plots and the bifurcated solutions are interpreted geometrically as critical points.

In \cref{sec:A3}, the third- and fourth-order expansion coefficients are derived explicitly as time integrals involving derivatives of the Lagrangian.  \cref{sec:num} presents four numerical examples for an inhomogeneous LJ potential and a homogeneous potential. These examples illustrate the analytical fold criterion derived in \cref{sec:sol}. Finally, \cref{sec:rem} summarizes the results and provides concluding remarks.

\section{Three-fold-type Bifurcation}
\label{sec:nes}

\subsection{Definition}\label{sec:def}
We consider a $T$-periodic solution $q(t)$, $q(t)=q(t+T)$, of the Euler--Lagrange equations
\begin{equation}\label{eq.motion}
	\frac{d}{dt} \frac{\partial L}{\partial \dot{q}} = \frac{\partial L}{\partial q}
\end{equation}
for the Lagrangian
\begin{equation}\label{eq.L}
	L = \frac{1}{2}|\dot{q}|^2 - U(q),
\end{equation}
where 
\begin{equation}\label{eq.U}
	U(q) = \sum_{1 \le j < i \le 3} u(r_{ij})
\end{equation}
is the total potential energy, and $u(r_{ij})$ is the two-body interaction potential as a function of the distance $r_{ij}$ between bodies $i$ and $j$.
The generalized coordinate $q$ is a nine-dimensional real vector,
\begin{equation}
	q = (q_1, q_2, \ldots, q_9) = (x_1, y_1, z_1, x_2, y_2, z_2, x_3, y_3, z_3),
\end{equation}
where $(x_i, y_i, z_i)$ is the position vector of body $i$.

A necessary condition for a bifurcation to occur from $q$ is that an eigenvalue $\kappa$ of the linear differential operator—the Hessian of the action functional \cite{fukuda2019},
\begin{equation}\label{eq.H}
	H(q) = -\frac{d^2}{dt^2} I_9 - \frac{\partial^2 U}{\partial q^2},
\end{equation}
crosses zero, where $I_9$ denotes the $9 \times 9$ identity matrix. Such an eigenvalue $\kappa$ is referred to as a critical eigenvalue.

Let $\phi = (\phi_1, \phi_2, \ldots, \phi_d)$ be a $9 \times d$ matrix whose columns are orthonormal real eigenfunctions spanning the eigenspace associated with the critical eigenvalue $\kappa$. 
Let $\psi = (\psi_1, \psi_2, \ldots)$ denote the remaining orthonormal real eigenfunctions.

Following the LS reduction, a perturbed orbit $q^{(b)}$ sufficiently close to the unbifurcated solution $q$ can be decomposed as
\begin{equation}\label{qb}
	q^{(b)} = q + \phi\bm{r} + \psi\bm{\epsilon},
\end{equation}
where $\bm{r}=(r_1,r_2,\ldots,r_d)^T$ is a $d$-dimensional real vector of representation variables and $\bm{\epsilon}=(\epsilon_1,\epsilon_2,\ldots)^T$ contains the real coefficients in the complementary space. The implicit function theorem determines $\bm{\epsilon}$ as a function of $\bm{r}$, with
\begin{equation}\label{epsilon}
	|\bm{\epsilon}|\to0
	\qquad\text{as}\qquad
	|\bm{r}|\to0.
\end{equation}
Consequently, the action functional
\begin{equation}
	S(f)=\int_0^T L(f,\dot f)\,dt
\end{equation}
for a $T$-periodic function $f(t)$ is reduced to a function of the representation variables $\bm{r}$:
\begin{equation}\label{Sr}
	S(\bm{r})
	=S\bigl(q^{(b)}\bigr)
	=S\bigl(q+\phi\bm{r}+\psi\bm{\epsilon}\bigr).
\end{equation}

Let $G(q)$ denote the finite spatiotemporal symmetry group of the unbifurcated solution $q$, and let $D(g)$ be the $d$-dimensional real orthogonal representation of $g\in G(q)$ on the critical eigenspace spanned by $\phi$.
When the representation group $D(q)=\{D(g)\mid g\in G(q)\}$, giving the symmetry of the reduced action
\begin{equation}\label{simS}
	S(\bm{r})=S\bigl(D(g)\bm{r}\bigr),
\end{equation}
is isomorphic to the cyclic group $C_3$ or the dihedral group $D_3$, the corresponding bifurcation is referred to as a three-fold-type bifurcation.

\subsection{Solutions up to the fourth order} 
\label{sec:sol}
For the three-fold-type bifurcation of the figure-eight choreography considered here, the critical eigenspace is two-dimensional, $d=2$ \cite{fukuda2025}. Under the restriction \eqref{simS}, $S(\bm{r})$ is expanded in polar coordinates, $\bm r=(r_1,r_2)^T=(r\cos\theta,r\sin\theta)^T$, as
\begin{equation}\label{S3}
	S(r,\theta)
	=S(q)
	+\frac{\kappa}{2}r^2
	+\frac{A_3}{3!}r^3\sin(3\theta)
	+\frac{A_4}{4!}r^4
	+O(r^5),
\end{equation}
where $A_3$ and $A_4$ are real expansion coefficients. In \eqref{S3}, the basis $(\phi_1, \phi_2)$ is assumed to be chosen such that $\phi_2$ possesses higher symmetry than $\phi_1$ when $D(q) \cong D_3$. For $D(q) \cong C_3$, there is no such preferred direction; hence, the phase $\varphi_3'$ in Eq.~(60) of \cite{fukuda2025}, which corresponds to a rotation of the basis $(\phi_1,\phi_2)$, is set to zero.

The corresponding critical-point equations are
\begin{align}
	\frac{\partial S}{\partial \theta} &= \frac{A_3}{2!}r^3\cos (3\theta) + O\left(r^5\right) = 0, \label{eq.dSdtheta} \\
	\frac{\partial S}{\partial r} &= \kappa r + \frac{A_3}{2!}r^2\sin (3\theta) + \frac{A_4}{3!}r^3 + O\left(r^4\right) = 0. \label{eq.dSdr}
\end{align}
In the subsequent analysis, we assume that $A_3\ne0$ and $A_4\ne0$.

Equation \eqref{eq.dSdtheta} implies that either $r = 0$ or 
\begin{equation}\label{eq.theta_sol}
	3\theta = \frac{\pi}{2} + m\pi \quad \text{for } m \in \mathbb{Z}.
\end{equation}
Since $r = 0$ trivially satisfies \eqref{eq.dSdr}, the origin $\bm{r} = \bm{0}$ always constitutes a trivial solution. 
By virtue of \eqref{qb} and \eqref{epsilon}, this trivial solution $r = 0$ corresponds exactly to the original unbifurcated solution $q$.

For non-trivial solutions with $r \neq 0$, substituting \eqref{eq.theta_sol} into \eqref{eq.dSdr} reduces it to the following quadratic equation in $r$:
\begin{equation}
	r^2 + (-1)^m \frac{3 A_3}{A_4} r + \frac{6\kappa}{A_4} = 0,
\end{equation}
which yields the two roots
\begin{equation}
	r = \frac{1}{2} \left( -(-1)^m \frac{3A_3}{A_4} \pm \sqrt{\left(\frac{3A_3}{A_4}\right)^2 - \frac{24\kappa}{A_4}} \right).
\end{equation}
By allowing a negative radius $r$ in the polar coordinate system, where $(-r, \theta)$ is identified with $(r, \theta + \pi)$, non-trivial solutions can be consistently expressed.
We find three solutions given by
\begin{equation}\label{sols}
	(r,\theta) = \left(r_-(\kappa), \theta_n\right) \quad \text{for } n = 1, 2, 3,
\end{equation}
and another three solutions given by
\begin{equation}\label{fsols}
	(r,\theta) = \left(r_+(\kappa), \theta_n\right) \quad \text{for } n = 1, 2, 3,
\end{equation}
where the radial functions $r_{\pm}(\kappa)$ are defined as
\begin{equation}\label{rpm}
	r_{\pm}(\kappa) = \frac{1}{2} \left( \left|\frac{3A_3}{A_4}\right| \pm \sqrt{\left(\frac{3A_3}{A_4}\right)^2 - \frac{24\kappa}{A_4}} \right),
\end{equation}
and the discrete angles $\theta_n$ are determined by:
\begin{equation}
	\theta_n = \begin{cases}
		\displaystyle \frac{\pi}{6} + \frac{2n\pi}{3} & \text{if } A_3 A_4 < 0, \\[1em]
		\displaystyle \frac{\pi}{6} + \frac{(2n+1)\pi}{3} & \text{if } A_3 A_4 > 0.
	\end{cases}
\end{equation}

The seven solutions given by \eqref{sols}, \eqref{fsols}, and $r = 0$ constitute all the critical points satisfying \eqref{eq.dSdtheta} and \eqref{eq.dSdr} up to the fourth order. Each set of three solutions, \eqref{sols} and \eqref{fsols}, differing by a $2\pi/3$ rotation, corresponds to the cyclic rotations of the bodies. Thus, the seven solutions represent three distinct solutions: two nontrivial solutions, each comprising three equivalent solutions, and one trivial solution.

When the critical eigenvalue $\kappa$ approaches zero at the bifurcation point, $r_-(\kappa)$ also vanishes according to \eqref{rpm}, independently of the sign of $\kappa$. Moreover, under our assumption that the eigenvalue crosses zero, $\kappa$ changes its sign at the bifurcation point. Therefore, the solutions \eqref{sols} exist on both sides of the bifurcation point in its immediate vicinity; they represent the branches that bifurcate directly from the original solution $q$. 

On the other hand, although the other solutions \eqref{fsols} also exist locally, they do not merge with $q$ as $\kappa \to 0$ because $r_+(\kappa) \to \left| 3A_3/A_4 \right| \neq 0$.
Consequently, they do not emerge directly from the bifurcation point, but instead collide with the primary branches to form a fold at $\kappa=\kappa_0$,
\begin{equation}\label{k0}
	\kappa_0 = \frac{3A_3^2}{8A_4},
\end{equation}
where the two sets of solutions \eqref{sols} and \eqref{fsols} coincide.

At this fold point $\kappa_0$, if $r_\pm(\kappa_0)=r_0$, given by
\begin{equation}\label{r0}
	r_0 = \left|\frac{3A_3}{2A_4}\right|,
\end{equation}
is sufficiently smaller than $1$, the lower-order truncation of the LS reduced action is well-justified, and the existence of these fold solutions is strongly anticipated in the actual system.

\subsection{Relative action and cusp structure}
\label{relative_action}
We define the relative action $\Delta S(r,\theta)$ as the difference between the action of a bifurcating solution and that of the original periodic orbit $q$:
\begin{equation}
	\Delta S(r,\theta) = S(r,\theta) - S(q).
\end{equation}
Substituting the solutions \eqref{sols} and \eqref{fsols} into \eqref{S3} yields the relative actions for the primary bifurcating solutions $(r_-(\kappa), \theta_n)$ and the fold solutions $(r_+(\kappa), \theta_n)$, respectively:
\begin{equation}\label{Sk}
	\Delta S\left(r_\pm(\kappa), \theta_n\right) = \frac{\kappa}{2}r_\pm^2(\kappa) - \frac{|A_3|}{3!} \frac{A_4}{|A_4|} r_\pm^3(\kappa) + \frac{A_4}{4!}r_\pm^4(\kappa) + O\left(r_\pm^5(\kappa)\right).
\end{equation}
Since the right-hand side of \eqref{Sk} is independent of the discrete angles $\theta_n$, we denote these actions simply as 
\begin{equation}\label{Skdef}
	\Delta S_\pm(\kappa)=\Delta S\left(r_\pm(\kappa), \theta_n\right).
\end{equation}

\begin{figure}[htbp]
	\centering
	\includegraphics[width=0.5\textwidth]{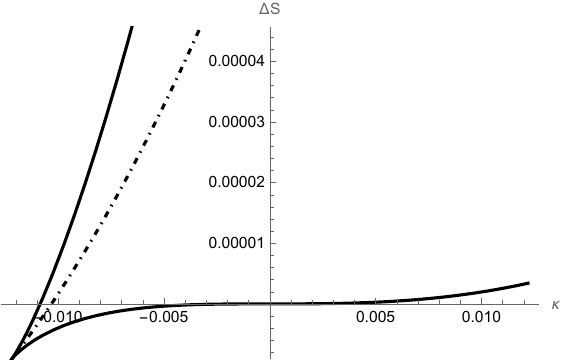}
	\caption{
		The relative actions $\Delta S_{\pm}(\kappa)$, defined in \eqref{Skdef}, evaluated for $A_3 = 0.51$ and $A_4 = -8.1$. 
		The solid curves represent $\Delta S_+(\kappa)$ and $\Delta S_-(\kappa)$, while the dot-dashed curve represents their average $\Delta \bar{S}(\kappa)$ given in \eqref{DeltaSbar}.
	}\label{fig:Sk}
\end{figure}

In Fig.~\ref{fig:Sk}, the relative actions $\Delta S_\pm(\kappa)$ evaluated for $A_3 = 0.51$ and $A_4 = -8.1$ are displayed as solid curves. The cusp located at $\kappa = \kappa_0 = -0.012$ corresponds to the fold point, where the two branches meet at $\Delta S_-(\kappa_0) = \Delta S_+(\kappa_0) = \Delta S_0$, with the critical action value given by
\begin{equation}\label{S0}
	\Delta S_0 = \frac{9 A_3^4}{128 A_4^3}.
\end{equation}
At the unbifurcated parameter value $\kappa=0$, we have $\Delta S_-(0) = 0$, whereas the other branch yields a non-zero residual action $\Delta S_+(0) = -9 A_3^4 / (8 A_4^3) \neq 0$.

By substituting the radial solutions \eqref{rpm} into \eqref{Sk}, the relative actions $\Delta S_\pm(\kappa)$ can be rearranged to explicitly reveal the cusp singularity:
\begin{equation}\label{cusp}
	\Delta S_\pm(\kappa) = \mp \frac{|A_3|}{6} \frac{A_4}{|A_4|} \left(\frac{6(\kappa_0-\kappa)}{A_4}\right)^{3/2} + \Delta \bar{S}(\kappa).
\end{equation}
The first term on the right-hand side of \eqref{cusp} exhibits the characteristic $3/2$-power fractional behavior of an ordinary cusp at $\kappa = \kappa_0$. The second term, which represents the average of the two action branches, is given by
\begin{equation}\label{DeltaSbar}
	\Delta \bar{S} (\kappa) = \Delta S_0 + \frac{r_0^2}{2} (\kappa-\kappa_0) - \frac{3}{2 A_4}(\kappa-\kappa_0)^2 + O\left(|\kappa-\kappa_0|^{5/2}\right).
\end{equation}
This term captures the smooth, regular background underlying the cusp structure. In Fig.~\ref{fig:Sk}, $\Delta \bar{S}(\kappa)$ is depicted by the dot-dashed curve originating from the fold point $\kappa = \kappa_0$.

\subsection{Graphical Explanation}
\label{sec:top}
\begin{figure}[h]%[htbp]
	\centering
	\begin{minipage}[t]{0.45\textwidth}
		\centering
		\includegraphics[width=\textwidth]{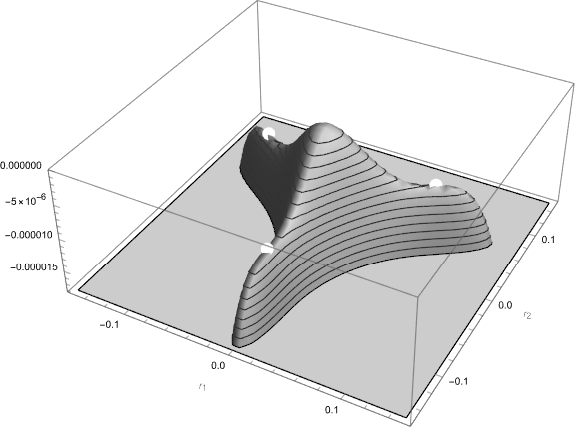}
		\vspace{0.3em}
		\\ (a)
	\end{minipage}
	\hfill
	\begin{minipage}[t]{0.45\textwidth}
		\centering
		\includegraphics[width=\textwidth]{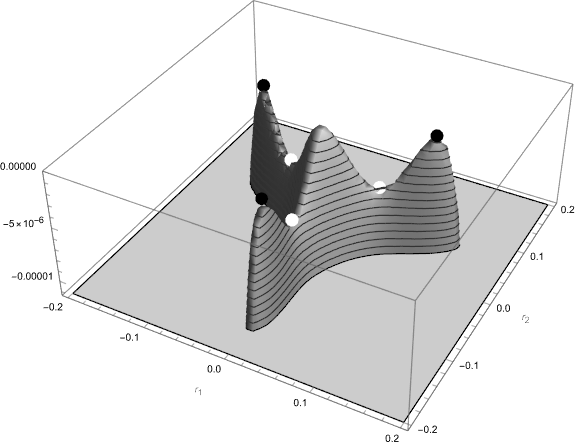}
		\vspace{0.3em}
		\\ (b)
	\end{minipage}
	
	\vspace{1.5em}
	
	\begin{minipage}[t]{0.45\textwidth}
		\centering
		\includegraphics[width=\textwidth]{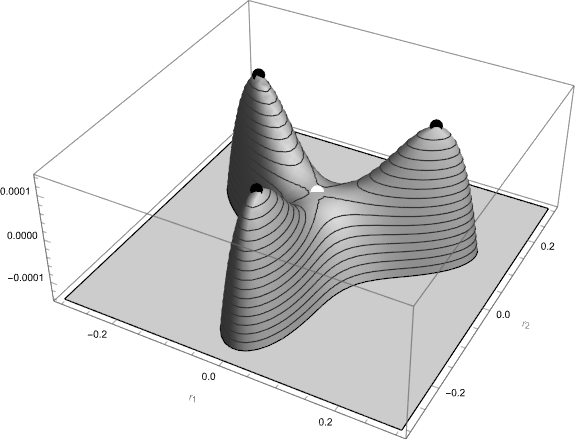}
		\vspace{0.3em}
		\\ (c)
	\end{minipage}
	\hfill
	\begin{minipage}[t]{0.45\textwidth}
		\centering
		\includegraphics[width=\textwidth]{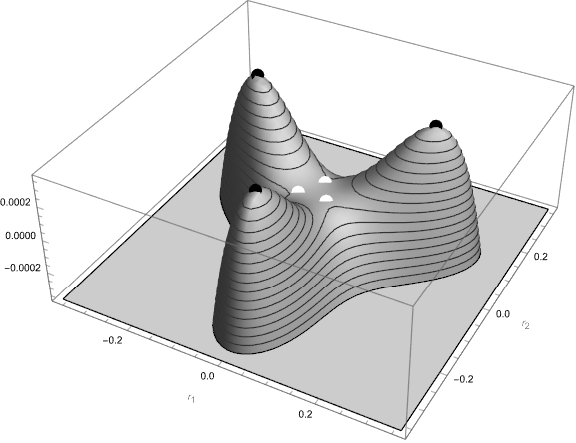}
		\vspace{0.3em}
		\\ (d)
	\end{minipage}
	
	\caption{
		Three-dimensional profiles of the relative action $\Delta S(r,\theta)$ for $A_3 = 0.51$ and $A_4 = -8.1$. 
		The horizontal axes represent $r_1 = r \cos\theta$ and $r_2 = r \sin\theta$. 
		(a) $\kappa = \kappa_0 = -0.012$, 
		(b) $\kappa = 0.9\kappa_0$, 
		(c) $\kappa = 0$, and 
		(d) $\kappa = -0.9\kappa_0$. 
		White spheres represent the primary bifurcating solutions \eqref{sols}, and black spheres represent the fold solutions \eqref{fsols}. The central extremum (peak or depression) corresponds to the original unbifurcated solution $q$.
	} \label{fig:c3D}
\end{figure}

In Fig.~\ref{fig:c3D}, three-dimensional landscapes of the relative action $\Delta S(r,\theta)$ over the $(r_1,r_2)$-plane are displayed for the same values of $A_3=0.51$ and $A_4=-8.1$ as in Fig.~\ref{fig:Sk}, where the fold occurs at $\kappa=\kappa_0=-0.012$. These plots provide a geometric visualization of the seven critical points obtained analytically in \eqref{sec:sol}.

In each panel of Figs.~\ref{fig:c3D}a--d, corresponding respectively to $\kappa=\kappa_0$, $\kappa_0<\kappa<0$, $\kappa=0$, and $\kappa>0$, the trivial solution $\bm r=\bm0$ is located at the origin. 
Depending on the sign of $\kappa$, this point appears as a central maximum or minimum. 
The three primary bifurcating solutions \eqref{sols} are shown by white spheres and the three fold solutions \eqref{fsols} are shown by black spheres. 

At the fold point shown in Fig.~\ref{fig:c3D}a, each primary bifurcating solution and its corresponding fold solution coalesce. Thus, the six nontrivial critical points reduce to three distinct critical points related by the three-fold symmetry.

For $\kappa_0<\kappa<0$, shown in Fig.~\ref{fig:c3D}b, the fold solutions represented by the black spheres remain at a finite distance from the origin. In contrast, the primary bifurcating solutions represented by the white spheres, located between the origin and the black spheres, approach the origin as $\kappa\to0$.

At the bifurcation point $\kappa=0$, shown in Fig.~\ref{fig:c3D}c, the three primary bifurcating solutions merge with the trivial solution at the origin.

For $\kappa>0$, shown in Fig.~\ref{fig:c3D}d, the primary bifurcating solutions, represented by white spheres, have passed through the origin and continue into the opposite sector relative to the fold solutions, represented by black spheres.

Thus, as $\kappa$ increases from $\kappa_0$, the six nontrivial critical points exhibit the following structure: three primary bifurcating solutions emerge from the fold at $\kappa=\kappa_0$, approach the origin, pass through the bifurcation point at $\kappa=0$, and continue into the opposite sector, while the three fold solutions remain at a finite distance from the origin.

\subsection{Integral expression of the coefficients}
\label{sec:A3}
The third-order expansion coefficient $A_3$ in \eqref{S3} is determined by expanding the action functional around the unbifurcated solution $q$ \cite{fujiwara2020}, yielding
\begin{equation}\label{A3}	
	A_3 = \begin{cases}
		A_3^{(0)}, & \text{if } D(q) \cong D_3, \\[1em]
		\displaystyle \frac{1}{3}\sqrt{\left(A_3^{(1)}\right)^2 + \left(3A_3^{(0)}\right)^2}, & \text{if } D(q) \cong C_3,
	\end{cases}
\end{equation}
where the component integrals $A_3^{(j)}$ are defined as
\begin{equation}\label{A3j}
	A_3^{(j)} = \binom{3}{j} \int_0^T \left[ \left(\phi_1(t) \frac{\partial}{\partial q}\right)^j \left(\phi_2(t) \frac{\partial}{\partial q}\right)^{3-j} L(q, \dot{q}) \right] \, dt.
\end{equation}

The time integrals in \eqref{A3j} can be evaluated numerically using computer algebra systems such as Mathematica \cite{mathematica}.
For a consistency check of the numerical integration in \eqref{A3j}, the following relation is useful:
\begin{equation}
	\kappa = \int_0^T \left[ \left(\phi_i(t)\frac{\partial}{\partial q}\right)^{2} L(q, \dot{q}) \right] \, dt \quad \text{for } i=1,2.
\end{equation}

On the other hand, the fourth-order coefficient $A_4$ originally involves a summation over all the complementary eigenfunctions $\psi_i$ of $H(q)$ associated with the non-zero eigenvalues $\lambda_i \neq 0$ at the bifurcation point:
\begin{equation}\label{A4}
	\begin{split}
		A_4 = {} & \int_0^T \left[ \left(\phi_2(t) \frac{\partial}{\partial q}\right)^{4} L(q, \dot{q}) \right] \, dt \\
		& - 3 \sum_{i} \frac{1}{\lambda_i} \int_0^T \left[ \left(\psi_i(t)\frac{\partial}{\partial q}\left(\phi_2(t)\frac{\partial}{\partial q}\right)^2\right)^2 L(q, \dot{q}) \right] \, dt.
	\end{split}
\end{equation}
Without further extensive analytical manipulation, \eqref{A4} cannot be evaluated numerically in a straightforward manner because it contains an infinite sum over these complementary eigenmodes.

\section{Numerical Calculations}
\label{sec:num}
In our numerical calculations for the figure-eight choreography published in \cite{fukuda2019, fukuda2025}, there are four three-fold-type bifurcations, as shown in Table~\ref{tab:res}.
\begin{table}[h]
	\centering
	\caption{Numerical results. Column $q$ distinguishes a figure-eight choreography with a period $T$. $C_y$ is a choreographic solution symmetric with respect to the $y$-axis, which bifurcates from $\alpha_+$ at $T=17.132$ \cite{fukuda2019}. $\infty$ is a homogeneous figure-eight choreography \cite{fukuda2019} with $T=2\pi$. The bifurcation parameters are the power of the potential $a$ for $\infty$, and the period $T$ for $\alpha_\pm$ and $C_y$. The fold point is the parameter value where the fold occurs. $\kappa_0$ and $\Delta S_0$ are the eigenvalue $\kappa$ and the action difference $\Delta S$ at this point, respectively. Numbers in parentheses denote the power of ten of the corresponding factors. $A_3$ and $A_4$ are the coefficients in \eqref{S3} calculated from $\kappa_0$ and $\Delta S_0$ via \eqref{k0} and \eqref{S0}. The column labeled (\ref{A3}) is the value of $A_3$ calculated by the integral expression \eqref{A3}. $r_0$ is the value of the representation variable $r$ at the fold point defined by \eqref{r0}. \label{tab:res}}
	\vspace{0.5em}
	\setlength{\tabcolsep}{4.5pt}
	\footnotesize
	\begin{tabular}{lcccccccc}
		\toprule
		$q$ & 
		\begin{tabular}{c} bifurcation \\ point \end{tabular} & 
		\begin{tabular}{c} fold \\ point \end{tabular} & 
		$\kappa_0$ & 
		$\Delta S_0$ & 
		$A_3$ & 
		\eqref{A3} & 
		$A_4$ & 
		$r_0$ \\
		\midrule
		$\alpha_+$ & $T=16.878$ & $16.875$ & $\phantom{-}6.6^{(-5)}$ & $\phantom{-}3.0^{(-9)}$ & $0.011$ & $0.011$ & $\phantom{-}0.74\phantom{0}$ & $0.023$ \\
		$\alpha_-$ & $T=14.836$ & $14.797$ & $-1.2^{(-2)}$          & $-9.1^{(-6)}$          & $0.51\phantom{0}$  & $0.45\phantom{0}$  & $-8.1\phantom{00}$           & $0.083$ \\
		$C_y$      & $T=17.235$ & $17.234$ & $-2.4^{(-4)}$          & $-5.1^{(-9)}$          & $0.059$ & $0.056$ & $-5.5\phantom{00}$           & $0.015$ \\
		$\infty$   & $a=0.9966$ & $\phantom{1}1.027$  & $\phantom{-}1.5^{(-2)}$ & $\phantom{-}3.8^{(-3)}$ & $0.036$ & $0.037$ & $\phantom{-}0.031$           & $1.7\phantom{00}$ \\
		\bottomrule
	\end{tabular}
\end{table}

The first column in Table~\ref{tab:res} indicates the original periodic solution $q$. The branches $\alpha_\pm$ in the second and third rows represent the figure-eight choreographies under the LJ potential:
\begin{equation}\label{LJ}
	u(r) = \frac{1}{r^{12}} - \frac{1}{r^6}.
\end{equation}
At any given period $T$, the action for $\alpha_+$ is higher than that for $\alpha_-$. These two branches coalesce at the minimum period $T = 14.479$, which corresponds to the saddle-node bifurcation point \cite{fukuda2019}.

The solution $C_y$ in the fourth row is a figure-eight-shaped choreography that bifurcates from the $\alpha_+$ branch at $T = 17.132$; it is characterized by being symmetric with respect to the $y$-axis but asymmetric with respect to the $x$-axis \cite{fukuda2019}.

The solution labeled $\infty$ in the fifth row corresponds to Moore's original figure-eight choreography \cite{moore} governed by the homogeneous potential:
\begin{equation}\label{homo}
	u(r) = -\frac{1}{a r^a},
\end{equation}
with a fixed period $T = 2\pi$.

For each $q$, the second column lists the bifurcation points alongside their respective bifurcation parameters, $T$ or $a$. The third column displays the fold point, which indicates the critical value of the bifurcation parameter where the fold of the bifurcated solution occurs.

The following columns labeled $\kappa_0$ and $\Delta S_0$ present their respective values, which are determined from the numerical plots of $\Delta S_\pm(\kappa)$ with $\kappa = \kappa(a)$ against $a$ and $\kappa = \kappa(T)$ against $T$, respectively.

Then, the values of $A_3$ and $A_4$ are given in the corresponding columns. These coefficients are determined by substituting the numerical data from the columns labeled $\kappa_0$ and $\Delta S_0$ into the algebraic relations derived from \eqref{k0} and \eqref{S0}:
\begin{align}
	A_3 &= 2 \sqrt{\frac{\kappa_0^3}{3\Delta S_0}}, \label{eq:A3_from_data} \\
	A_4 &= \frac{\kappa_0^2}{2\Delta S_0}. \label{eq:A4_from_data}
\end{align}
For comparison, the column labeled \eqref{A3} presents the values of $A_3$ directly evaluated via the numerical integration of \eqref{A3} combined with \eqref{A3j}.

Here, for the $\alpha_-$ branch in the third row, a 12\% discrepancy in $A_3$ is observed between the values derived from the algebraic fitting in the column labeled $A_3$ and the direct numerical integration of \eqref{A3j} in the column labeled \eqref{A3}, whereas the discrepancies for the other three examples are remarkably smaller. 
This localized discrepancy arises because the numerical pinpointing of the fold point $(\kappa_0, \Delta S_0)$ from the plot of $\Delta S_\pm(\kappa)$ with $\kappa=\kappa(T)$ against $T$ is more sensitive for this case than for the other three cases.

\begin{figure}[h]%[htbp]
	\centering
	\begin{minipage}[t]{0.45\textwidth}
		\centering
		\includegraphics[width=\textwidth]{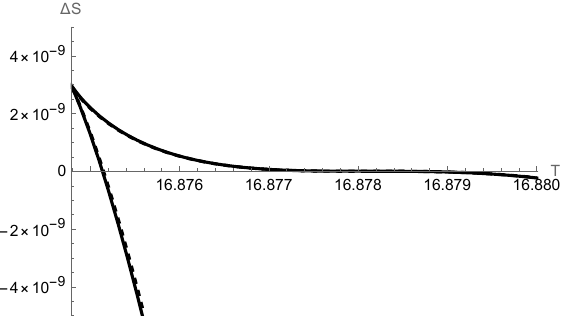}
		\vspace{0.3em}
		\\ (a)
	\end{minipage}
	\hfill
	\begin{minipage}[t]{0.45\textwidth}
		\centering
		\includegraphics[width=\textwidth]{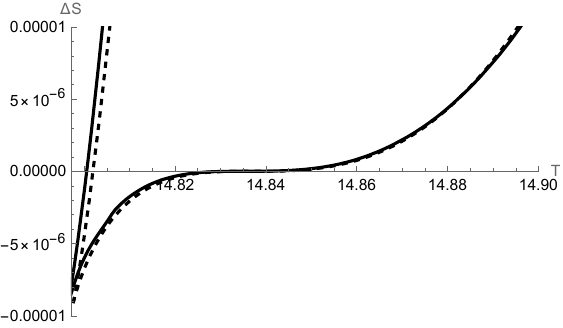}
		\vspace{0.3em}
		\\ (b)
	\end{minipage}
	
	\vspace{1.5em}
	
	\begin{minipage}[t]{0.45\textwidth}
		\centering
		\includegraphics[width=\textwidth]{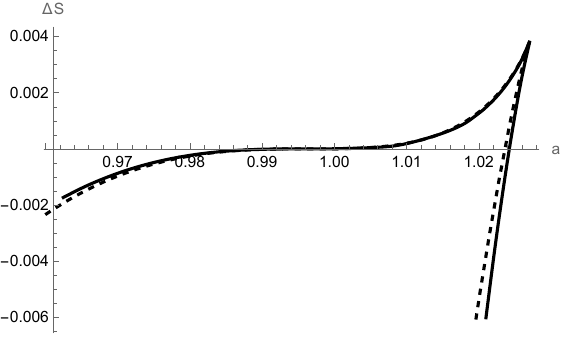}
		\vspace{0.3em}
		\\ (c)
	\end{minipage}
	\hfill
	\begin{minipage}[t]{0.45\textwidth}
		\centering
		\includegraphics[width=\textwidth]{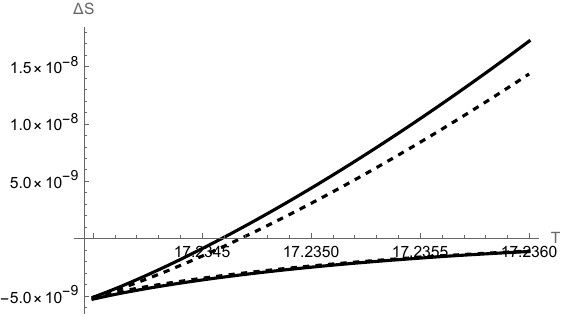}
		\vspace{0.3em}
		\\ (d)
	\end{minipage}
	
	\caption{Plots of $\Delta S_\pm(\kappa(T))$ and $\Delta S_\pm(\kappa(a))$, defined in \eqref{Skdef}, for the four bifurcations:
		(a) Bifurcation from $\alpha_+$ at $T=16.878$ under the LJ potential \eqref{LJ}, with $\kappa(T)=0.336 - 0.0199 T$. 
		(b) Bifurcation from $\alpha_-$ at $T=14.836$ under the LJ potential \eqref{LJ}, with $\kappa(T)=-142.491 + 18.9229 T - 0.628098 T^2$.
		(c) Bifurcation from $C_y$ at $T=17.235$, which itself bifurcated from $\alpha_+$ at $T=17.132$ \cite{fukuda2019}, under the LJ potential \eqref{LJ}, with $\kappa(T)=-0.671 + 0.0389 T$.
		(d) Bifurcation from the homogeneous figure-eight choreography at $a=0.9966$ under the homogeneous potential \eqref{homo}, with $\kappa(a)=-0.504 + 0.506 a$.
		In (a)--(d), the solid curves represent the numerically exact bifurcated solutions, while the dashed curves represent the theoretical predictions obtained from \eqref{Sk} using the coefficients $(A_3, A_4)$ listed in Table~\ref{tab:res} alongside the expansions of $\kappa(T)$ or $\kappa(a)$ around $\kappa=0$ given above.
	}
	\label{fig16878Sk}
\end{figure}

The last column of Table~\ref{tab:res} lists the values of $r_0$ defined in \eqref{r0}. With the exception of the fourth example designated as $\infty$, the remaining three cases satisfy the condition $r_0 \ll 1$. 
This indicates that the fourth-order truncation is expected to provide a reliable local description of the fold behavior in these three cases. Indeed, as plotted in Figs.~\ref{fig16878Sk}a--c, the dashed curves of $\Delta S_\pm\left(\kappa(T)\right)$ calculated from \eqref{Sk} using the fitted coefficients $(A_3, A_4)$ and $\kappa(T)$ specified in Table~\ref{tab:res} exhibit excellent agreement with the solid curves computed directly from the numerically exact bifurcated solutions even away from the fitted fold point.

On the other hand, for the fourth example designated as $\infty$, a unique behavior is observed. As illustrated in Fig.~\ref{fig16878Sk}d, even though $r_0 = 1.719 > 1$, the theoretical prediction of $\Delta S_\pm\left(\kappa(a)\right)$ obtained from \eqref{Sk} as a dashed curve tracks the solid curve of the numerically exact bifurcated solutions remarkably well; here, the prediction utilizes $\kappa(a)$ provided in its caption.

\section{Summary and Discussions}
\label{sec:rem}
We have demonstrated that for three-fold-type bifurcations, the bifurcated branches inevitably undergo a fold transition as a function of the bifurcation parameter at $\kappa=\kappa_0$ on the side of $\kappa A_4 > 0$, provided that $r_0\ll 1$.

It was previously shown that among the bifurcations from the figure-eight choreography and its bifurcation cascade, the three-fold-type bifurcation is characterized by the emergence of less symmetric bifurcated solutions on both sides of the bifurcation point, $\kappa>0$ and $\kappa<0$ \cite{fukuda2025}.
Therefore, for the figure-eight choreography and its bifurcation cascade, whenever less symmetric bifurcated solutions emerge on both sides of the bifurcation point, these branches necessarily undergo a fold transition on the side where $\kappa A_4 > 0$, provided that $r_0\ll 1$.

It should be noted, however, that the evaluation of $r_0$ inherently requires a prior determination of the fold point itself, because the explicit integral expression for $A_4$ given in \eqref{A4} remains computationally intractable via direct numerical integration. While this requirement marks a current methodological limitation, both the agreement of the $A_3$ values obtained via \eqref{A3} and \eqref{eq:A3_from_data}, and the close agreement between our algebraic fitting and the exact numerical solutions even away from the fold point, robustly validate the geometric and topological framework presented herein.

Regarding the condition $r_0 \ll 1$, the fourth example designated as $\infty$ among the four numerical examples is particularly noteworthy. There, this condition is clearly violated as $r_0 = 1.7 > 1$. Nevertheless, a fold transition still manifests.
As visually demonstrated through the topological landscapes in \cref{sec:top}, such a fold appears to be a topological necessity; the formation of specific auxiliary crests representing the folded solutions seems indispensable for generating the required saddle points in close proximity to the primary peak that represents the original choreography $q$.

In the three-body problem, a fold point typically constitutes a bifurcation point in its own right, commonly referred to as a fold, or saddle-node, bifurcation. Within a fold bifurcation, the spatiotemporal symmetry of the bifurcated branch is strictly preserved, whereas other qualitative characteristics undergo a distinct transition. Consequently, through this fold phenomenon, the three-fold-type bifurcation effectively yields two coexisting solutions that possess fundamentally different physical behaviors.

For example, this transition appears to be manifested at the fold point $a = 1.027$ for the branch bifurcating from $a = 0.9966$ under the homogeneous potential \eqref{homo}.
Beyond the fold point, the qualitative character of the bifurcated solution changes significantly; specifically, as $a \to 1$, the solution trajectory after the fold asymptotically resembles a brake orbit \cite{brake}.

Although the classification of bifurcations via the LS reduction is well-established \cite{Golubitsky2, Poston}, its explicit application to a highly nonlinear Lagrangian system like the figure-eight choreography has remained unexplored. 
Having established this explicit application, the fold phenomenon is expected to manifest in the general few-body problem, provided that the bifurcation dynamics can be rigorously described by the LS reduced action in two dimensions with three-fold symmetry.

\end{document}